\documentclass[12pt]{iopart}
\usepackage{iopams}  
\usepackage{graphicx}
\begin{document}

\title[Time-dependent two-level models and zero-area pulses]{Time-dependent two-level models and zero-area pulses}

\author{J M S Lehto and K-A Suominen}

\address{Department of Physics and Astronomy, Turku Centre for Quantum Physics and Laboratory of Quantum Optics, University of Turku, FI-20014 Turku, Finland}
\ead{jaakko.lehto@utu.fi}
\begin{abstract}
Time-dependent two-level models have been an important element in physics, and in particular in quantum optics since 1930's. We review the basics of these models and focus on the dynamics induced by off-resonant zero-area pulses by using a Sech-Tanh pulse model as an example. We show that the final transition probability for this model is described accurately by the Dykhne-Davis-Pechukas approach for certain parameter regions. Finally, we note the potential of such zero area pulse models in quantum control and quantum information.
\end{abstract}

\maketitle

\section{Introduction}

The dynamics of two-level systems, either induced or tailored with time-dependent external fields, have for many decades been the true workhorse for quantum optics, atomic and molecular physics~\cite{AllenEberly1975,Shore2011,Nakamura2012} and more recently, for quantum information and quantum computing~\cite{Stenholm2005,Nori2010}. Due to the exponentially increasing interest in the latter two topics, various aspects originally studied for the purposes of atomic physics are now re-examined and developed further in the context of solid state physics. The key element in many cases is the interaction of matter with electromagnetic fields, especially light.

We can crudely divide the dynamics of two-level systems into three categories: crossing of energy levels under some interaction between them, coupling of levels with external pulses, or the combination of both cases, which can be achieved e.g. by using chirped pulses. Despite their simple appearance, very few two-level models with explicit time-dependence in their Hamiltonian can be solved analytically. 

The Landau-Zener model~\cite{Landau1932,Zener1932}, for which credit should also be given appropriately to St\"uckelberg~\cite{Stuckelberg1932} and Majorana~\cite{Majorana1932}, describes the simplest possible level crossing case, with linearly changing energy separation and constant coupling. Since in most cases we can approximate any time-dependence near the crossing point with linear behaviour, and any coupling as constant, this model has had tremendous use in physics since 1932. 

The other important model is the Rosen-Zener model, for which we have constant energies and a hyperbolic secant pulse that couples the two levels~\cite{RosenZener1932}. In general, the resonant case can be solved for any pulse form with the area theorem (as described later in sec.~\ref{areatheorem})~\cite{AllenEberly1975}, but solutions to the off-resonant case are rare, making the Rosen-Zener model a strong representative of such pulsed models, in addition to the trivially solved square pulse case (Rabi model). In 1969, Demkov and Kunike provided an analytic solution to a representative of the chirped pulse case~\cite{DemkovKunike1969}, with a hyperbolic tangent function as the energy difference, and a hyperbolic secant pulse to couple the levels. The model is limited to both functions having the same time scale, but although the separate time scale case has been studied~\cite{Suominen1992a}, so far there is no analytic solution. 

Perturbative approaches to time-dependent models address typically two different regions of dynamics. In the case of weak coupling one can use the basic perturbative approaches~\cite{Light1977,Rodgers1987}, but in the case of slow dynamics, often related to strong coupling or semiclassical limit, one can make studies in the adiabatic basis, which corresponds to the instantaneous eigenstates of the system Hamiltonian, also called adiabatic states, see sec.~\ref{adiabatic}. Then one seeks nonadiabatic corrections that appear as transitions between these adiabatic states~\cite{Nakamura2012,Suominen1991,Lehto2015}. A specific tool for studying the nonadiabatic transitions was developed in 1960's by Dykhne~\cite{Dykhne1960,Dykhne1962}, and David and Pechukas~\cite{DavisPechukas1976}. This DDP approach relies on finding the complex zeros of the adiabatic energy difference. Interestingly, it recovers the exact Landau-Zener result for all parameter regions, and its extensions provide also the exact solutions to the Rosen-Zener model and the Demkov-Kunike model beyond the adiabatic limit~\cite{Suominen1992a,Suominen1992c}. 

Pulses with zero areas have been originally studied in the context of self-induced transparency, since a small area pulse becomes a zero area pulse after propagating a while in a resonant medium, see the excellent discussion in Ref.~\cite{Shore2011} and references therein. Zero area means that an integral over the pulse amplitude reaches a null value by the end of the pulse, implying typically that the pulse amplitude is an odd function in time. Basically, when purely antisymmetric, such pulses are expected to do nothing in the resonant case as the second part of the pulse cancels any excitation created by the first part. However, for off-resonant cases such pulses do have an effect. As such pulse models then become yet another tool for the case of quantum control, we find that to know their properties is a useful topic for research. They also provide an interesting case for demonstrating and testing the DDP approach for parameter regions that do not correspond to the adiabatic limit.

In this paper we first review general concepts about two-level systems with explicit time-dependence in the Hamiltonian, with focus on zero area pulses. Then we demonstrate the usefulness of the DDP approach with the Sech-Tanh model that has been presented recently~\cite{VasilevVitanov2006}. Although we can not yet use the DDP results for uncovering the full solution for all parameter regions, we find that when the detuning is large enough we get a quantitatively good agreement, and for other parameter regions an agreeable qualitative description.

\section{Mathematical formalism}

\subsection{Schr\"{o}dinger equation}

We consider the dynamics of two-level quantum systems subject to external driving. We assume the dynamics to be coherent, so the systems obey the time dependent Schr\"{o}dinger equation,
\begin{equation}
\rmi \partial_{t}\bi{\psi}(t) = H(t)\bi{\psi}(t),
\label{eqn:scheq}
\end{equation}
written in units where $\hbar = 1$. The interaction between the system and the external fields leads to the time dependencies in the Hamiltonian which is given in the so-called diabatic basis consisting of the time independent states of the non-interacting system as
\begin{equation}
H(t)  = \left( \begin{array}{lr}
			\varepsilon (t) & V(t) \\
			V(t) & -\varepsilon (t)
					\end{array}\right). 
\label{eqn:hamiltonian}		 		
\end{equation}
The state is given by the normalized complex vector $ \bi{\psi} (t) = \left( c_{2}(t), c_{1}(t) \right)^{T} $, where $c_{1, 2}$ are called the probability amplitudes of the corresponding basis states. In quantum optics, this model describes coherent population transfer induced by the semiclassical laser field between the two levels in the dipole and rotating-wave (RWA) approximations. Then the difference of the diabatic energy levels, $2\varepsilon(t)$, is called detuning and the function proportional to the coupling, $2V(t)$, is the Rabi frequency. 

In general, we assume that system is initially in state one, i.e., $c_{1} = 1$ and we want to find out what is the population of the other state at some later time. This corresponds to the transition probability $P(t)$ and is given by $P(t) = \vert c_{2}(t)\vert^{2}$. The final transition probability is denoted simply by $P$.

\subsection{Adiabatic basis}\label{adiabatic}

Another important concept in quantum dynamics is the adiabatic basis which is formed by the eigenstates of the diabatic Hamiltonian. For the Hamiltonian (\ref{eqn:hamiltonian}) they are given by 
\begin{equation}
\chi_{+}(t) = \pm \left( \begin{array}{c} \cos ( \frac{\theta(t)}{2} ) \\
									  \sin	( \frac{\theta(t)}{2}) \end{array} \right) , \qquad \chi_{-}(t) = \pm \left( \begin{array}{c} -\sin ( \frac{\theta(t)}{2}) \\
										   \cos	( \frac{\theta(t)}{2} ) \end{array} \right),
\label{eqn:chi}										   
\end{equation}
where $V/\varepsilon = \tan (\theta)$. A convenient and physically motivated assumption is that $V(t_{i}) = V(t_{f}) = 0$, so that the diabatic and adiabatic bases coincide in the initial and final times which makes the comparison of the transition probabilities straightforward in these bases. The corresponding eigenvalues, the quasienergies, are given by  
\begin{eqnarray}
E_{\pm}(t) &= \pm \rho (t)\\ \nonumber
		   &= \pm \sqrt{\varepsilon^{2}(t) + V^{2}(t)}.
\label{eqn:eigenenergies}		  
\end{eqnarray}
In this basis, the Hamiltonian reads
\begin{equation}
H_{a}(t) = \left(\begin{array}{lr} \rho(t) & \rmi \gamma (t) \\
								-\rmi \gamma (t) & -\rho(t)	\end{array}\right),
\label{eqn:adiabH}								
\end{equation}
where the adiabatic coupling $\gamma (t)$ is defined by 
\begin{eqnarray}
\gamma(t) &\equiv -\langle \chi_{+} \vert \dot{\chi}_{-}(\tau) \rangle \nonumber \\ 
&= \frac{\varepsilon(t)\dot{V}(t) - \dot{\varepsilon}(t)V(t)}{2\left(\varepsilon^{2}(t) + V^{2}(t) \right)} \nonumber \\
&= \frac{\dot{\theta}(t)}{2}, 
\label{eqn:acoupl}
\end{eqnarray}
and the overhead dot stands for time derivation. This coupling vanishes in the adiabatic limit, i.e., when the Hamiltonian changes slowly, and if the system is initially in an eigenstate, it stays there at all times during the adiabatic evolution. This requires also that the adiabatic energies do not cross. Diabatic levels, on the other hand, can cross, i.e. the detuning can go to zero, and the diabatic basis states can therefore change their character during the evolution; the state initially lower in energy can become the excited state. This is used in rapid adiabatic passage, where one obtains a robust complete population transfer (CPT) between the diabatic states by an adiabatic evolution. The condition for adiabatic evolution is given by 
\begin{equation}
\vert \varepsilon(t)\dot{V}(t) - \dot{\varepsilon}(t)V(t) \vert \ll \left[ \varepsilon^{2}(t) + V^{2}(t) \right]^{3/2}.
\label{eqn:adiabaticitycondition}
\end{equation}
Although the evolution is adiabatic, it is still considered fast in comparison to the decoherence and dissipation time scales of the physical system, hence the name rapid adiabatic passage.

\subsection{Area theorem}\label{areatheorem}

In a resonant situation, $\varepsilon \equiv 0$ and equation (\ref{eqn:scheq}) can be solved. With the above initial condition, the transition probability is simply  
\begin{equation}
P(t) = \sin^{2}\left[\int_{t_{i}}^{t}V(x)\rmd x\right],
\end{equation}
which shows that the transition probability at time $t$ depends only on the total pulse area,
\begin{equation}
A\left(t, \, t_{i}\right) = 2\int_{t_{i}}^{t}V(x)\rmd x,
\end{equation}
accumulated at this time and not on the shape and other details of the pulse. The factor of $2$ in $A$ arises from our choice of the Hamiltonian (\ref{eqn:hamiltonian}). We denote $A \equiv A\left(t_{f}, \, t_{i}\right) $ for the total pulse area of the process. It should be noted that this way $P(t)$ can also obtain any value between $0$ and $1$. For example, for constant Rabi frequency, $P(t)$ oscillates sinusoidally. These Rabi flops are also present in off-resonance but the maximum population transfer diminishes and the flopping frequency grows. 

The area theorem allows one to design resonant pulses which produce the exact amount of coherent excitation that is needed in a process. In particular, $A= \pi\left(2k +1\right)$ gives a complete population transfer (CPT) with any integer $k$, $A = 2\pi k$ a complete population return (CPR) and with $A = \pi\left(k + \frac{1}{2}\right)$ one obtains an equal superposition of the two basis states. The latter case is very useful for quantum information purposes~\cite{Stenholm2005}. However, the parameters of the pulse must be controlled very precisely in order to obtain the exact pulse area and for it to then produce the excitation that is sought for, so this scheme is not very robust. The area theorem and such concepts as $\pi/2$ and $\pi$ pulses are extensively used in NMR physics, and more recently in quantum computing studies.

Partly because of the area theorem, general interest in zero area pulse models has been limited. It has been shown relatively recently, however, that by going off the resonance, even CPT can be obtained in a quite robust manner \cite{VasilevVitanov2006}.

\subsection{Symmetry considerations}

We are interested in the models with zero pulse area and therefore imposed the antisymmetry condition $V(-t) = -V(t)$ on the diabatic coupling 
which automatically takes care of the pulse area. With this simplifying condition, the obvious choices for the detuning would similarly be those that are either odd or even with respect to time. 

The case where also the detuning is an odd function seems like an interesting special case initially. Although at the time of the level-crossing, $t= 0$, the coupling also vanishes, one can expect that there would be transitions between the diabatic states because of the finite time of the transition. And indeed, there generally is some transient population of the state not populated initially. It turns out, however, that simple symmetry considerations show the final transition probability to be strictly zero \cite{VitanovKnight1995}. This result is actually valid for any $N$-level system. 

Therefore we concentrate our efforts to the case with even detuning and from now on we impose the condition $\varepsilon(-t) = \varepsilon(t)$. One further note on these symmetries: It is easy to see from the equations (\ref{eqn:eigenenergies}) and (\ref{eqn:acoupl}) that in the adiabatic basis the odd $V$ and even $\varepsilon$ combination (or, more generally, when $\varepsilon$ and $V$ are of different parity) will give adiabatic coupling and energies that are both even functions of time, $\gamma(-t) = \gamma(t)$ and $\rho(-t) = \rho(t)$. 

\section{Models}

\subsection{Rosen-Zener model}

The Rosen-Zener model was originally introduced to describe the double Stern-Gerlach experiment \cite{RosenZener1932} and it consists of
\begin{equation}
\varepsilon(t) = a, \qquad V(t) = b\, \textrm{sech}\left(\frac{t}{T}\right),
\end{equation}
where $a$, $b$ and $T$ are positive constants. Obviously, this model does not have a pulse of zero area but a symmetric one instead. Nevertheless, it serves as a good reference model to understand the effects of asymmetries of the pulses to dynamics of the system. It can be exactly solved and the final excitation probability is given by 
\begin{equation}
P = \sin^{2}\left(\pi b T\right)\textrm{sech}^{2}\left(\pi a T\right).
\label{eqn:Prz}
\end{equation}
It is remarkable that the transition probability factorizes into two terms. First term is in accordance with the area theorem since the pulse area is exactly $A =2 b\pi T $. The other term controls the maximum of $P$. It is non-oscillating and depends only on the detuning. However, it is also well-known that the factorization property of (\ref{eqn:Prz}) does not hold in general, contrary to the original conjecture of Rosen and Zener \cite{RosenZener1932}. Furthermore, it has been shown for models with purely positive pulse area, that by going to the off-resonant case the asymmetry of the pulse has the consequence of eliminating the zeros of $P$ \cite{BambiniBerman1981, Robinson1981}. 
\begin{figure}[h]
\begin{center}
\includegraphics[scale=0.6]{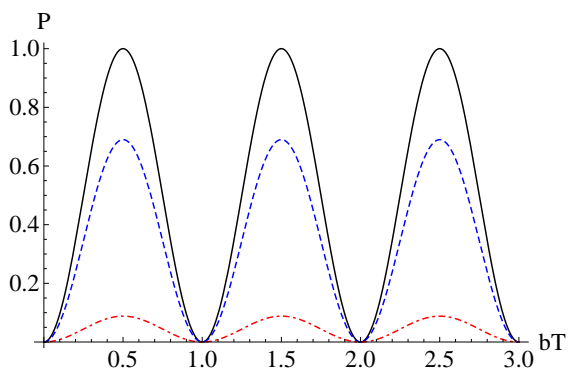} 
\caption{Transition probability for Rosen-zener model as a function of the coupling constant $bT$ for values $aT = 0$ (black), $aT = 0.1$ (blue and dashed) and $aT = 0.3$ (red and dotdashed).}
\label{fig:rzmodel1d}
\end{center}
\end{figure}

\subsection{Sech-Tanh model}
\label{subsec:ST}

\begin{figure}[h]
\begin{center}
\includegraphics[scale=0.7]{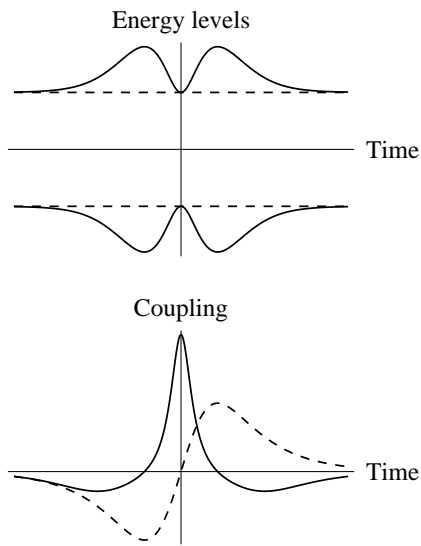} 
\caption{Schematic picture for energy level structure and coupling in diabatic (dashed line) and adiabatic (full line) bases for the Sech-Tanh model.} 
\label{fig:STscheme}
\end{center}
\end{figure}

A slight variation of the Rosen-Zener model with zero pulse area is the  Sech-Tanh model~\cite{VasilevVitanov2006}, given by (see figure \ref{fig:STscheme})
\begin{equation}
\varepsilon(t) = a, \qquad V(t) = b\,\textrm{sech}\left(\frac{t}{T}\right)\tanh\left(\frac{t}{T}\right).
\end{equation}
We scale the time to be dimensionless, $\tau = t/T $, and get the new parameters as $a \hookrightarrow \tilde{a} = aT$ and $b \hookrightarrow \tilde{b} = bT$.

Actually, this model has a pulse of zero area, albeit a symmetric one, also in the adiabatic basis. The diabatic levels do not cross, so the final transition probability $P$ is the same in both bases. Because of the usual association of the adiabatic limit with the strong coupling one, one would expect $P$ to vanish with large $b/a$. However, the matter is more complicated as is obvious from the scaling to the dimensionless variables above. The adiabatic limit $T\rightarrow \infty$ would correspond to both $\tilde{a}\rightarrow \infty$ and $\tilde{b}\rightarrow \infty$ but in this process the ratio $\tilde{a}/\tilde{b} \equiv a/b$  remains constant. In fact, the final transition probability tends to unity in the limit $b/a \rightarrow \infty$, as demonstrated in \cite{VasilevVitanov2006}. There this CPT mechanism was explained by the behavior of the nonadiabatic coupling. 

\begin{figure}[h]
\begin{center}
\includegraphics[scale=0.6]{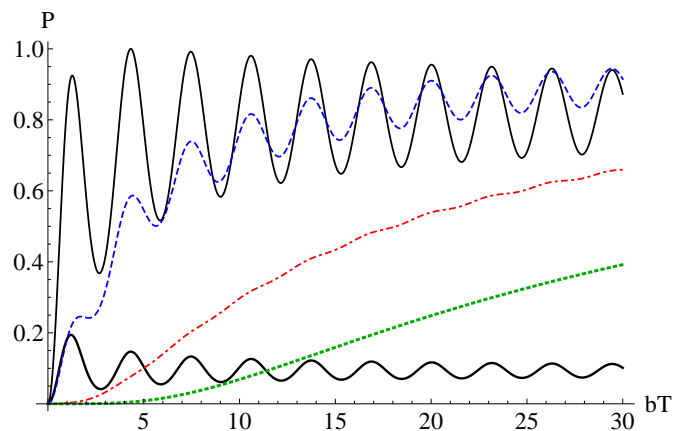} 
\caption{Transition probability for the Sech-Tanh model as a function of the coupling constant $bT$ for values $aT = 0.1$ (thick black line), $aT = 0.5$ (black solid line), $aT = 1$ (blue and dashed), $aT = 2$ (red and dotdashed) and $aT = 3$ (green and dotted).}
\label{fig:sechtanhmodel1d}
\end{center}
\end{figure}

When $b/a \rightarrow \infty$, the positive part of $\gamma(\tau)$, located in the interval $\left[-\tau_{0}, \tau_{0}\right]$, where $\tau_{0} = \mathrm{arcsinh}\left(1\right)$, behaves like a delta function. At the same time, the energy splitting is minimal and dependent only on $a$ because of $E_{\pm}(0) = \pm a$. Therefore, the area theorem suggests 
\begin{eqnarray}
P &\approx \sin^{2}\left[\int_{-\tau_{0}}^{\tau_{0}}\rmd x \gamma(x)\right], \; b/a \gg 1 \\ \nonumber
&= \sin^{2}\left[\theta(\tau_{0})\right],
\label{eqn:pcpt}
\end{eqnarray}
where the argument has the value 
\begin{equation}
\theta(\tau_{0}) = \arctan\left[b/(2a)\right].
\end{equation}
It follows that $P \rightarrow 1$ as $b/a \rightarrow \infty$. The robustness of this scenario is ensured by restricting the detuning not to be too small $(aT \gg 1/\sqrt{2})$ in order to prohibit the transitions happening due to the tails of the pulse. All of the features described here are also evident in figure \ref{fig:sechtanhmodel1d} showing the behavior of numerically obtained $P$ with different parameters. 

In reference \cite{VasilevVitanov2006} the focus was to reveal the surprising CPT phenomenon and the authors were interested on the equation (\ref{eqn:pcpt}) only in the limit $b/a \rightarrow \infty$. The exact solution for this model is not known. In the following section we will discuss a way of obtaining a more accurate analytic approximation for $P$ that works also outside the strong-coupling region. We use the DDP theory, and show how it can be used to explain the above-mentioned features of the model.

\section{Dykhne-Davis-Pechukas formula}

Dykhne-Davis-Pechukas (DDP) formula provides an useful and often very accurate approximation for the transition probability. It is asymptotically exact in the adiabatic limit \cite{Dykhne1960, Dykhne1962, DavisPechukas1976}. In this limit the effect of adiabatic coupling to the transition probability becomes universal, i.e., independent of the model and the transition probability is determined completely by the analytically continued eigenenergies. The original DDP formula reads
\begin{equation}
P = \exp \left(-2 \mathrm{Im}D(t_{c}) \right),
\label{eqn:ddp}
\end{equation}
where
\begin{equation}
D(t) = 2 \int_{0}^{t}\sqrt{\rho(s)}\mathrm{ds}.
\label{eqn:ddpintegral}
\end{equation}

In the case of multiple zero points, one should include the contribution from each zero point $t_{c}^{k}$ to the transition amplitude and the generalization of the DDP formula reads~\cite{Vitanov1999}
\begin{equation}
P_{DDP} = \left\vert \sum_{k = 1}^{N} \Gamma_{k} e^{i D\left(t_{c}^{k}\right)} \right\vert^{2}, 
\label{eqn:genDDP}
\end{equation}
where
\begin{equation}
\Gamma_{k} = 4 i \lim_{t \rightarrow t_{c}^{k}} \left( t - t_{c}^{k}\right) \gamma(t),
\label{eqn:isogamma}
\end{equation}
and $\gamma$ is the adiabatic coupling. Usually $\Gamma_{k} = \pm 1$ and this is the case also for Sech-Tanh model.

\subsection{Zero point structure for Sech-Tanh model}

For the Sech-Tanh model, the complex transition points $\tau_{c}^{(k)}$ are given by
\begin{equation}
a^{2} +b^{2}\mathrm{sech}^{2}(\tau_{c})\tanh^{2}(\tau_{c}) = 0.
\end{equation}
Using the formula $\tanh^{2}(x) + \mathrm{sech}^{2}(x) = 1$, we get 
\begin{equation}
a^{2} + b^{2}\mathrm{sech}^{2}(\tau_{c}) - b^{2}\mathrm{sech}^{4}(\tau_{c}) = 0.
\end{equation}
This gives
\begin{eqnarray}
\cosh^{2}(\tau_{c}) &= -\frac{1}{2}\left(b/a\right)^{2} \pm \frac{1}{2}\sqrt{\left(b/a\right)^{4} + 4 \left(b/a\right)^{2}},\\
&\equiv X_{\pm},
\end{eqnarray}
Furthermore, using the identity $\cosh(2x) + 1 = 2\cosh^{2}(x)$, we finally get
\begin{eqnarray}
\tau_{c}^{(k)} &= \frac{1}{2}\mathrm{Arcosh}\left(2X_{\pm} - 1 \right)\\
		  &= \pm \frac{1}{2}\mathrm{arcosh}\left(2X_{\pm} - 1 \right) + \dot{\imath}\pi k, 
		  \label{eqn:complextransitionpoint}
\end{eqnarray}
where $k$ is an integer ($\mathrm{Arcosh}$ written with capital is the principal value). It should be noted that the position of zero points depends only on the combination $b/a$ and, by going to the original variables, $T$ is only an overall multiplying factor. 

\begin{figure}[hb]
\begin{center}
\includegraphics[scale=0.65]{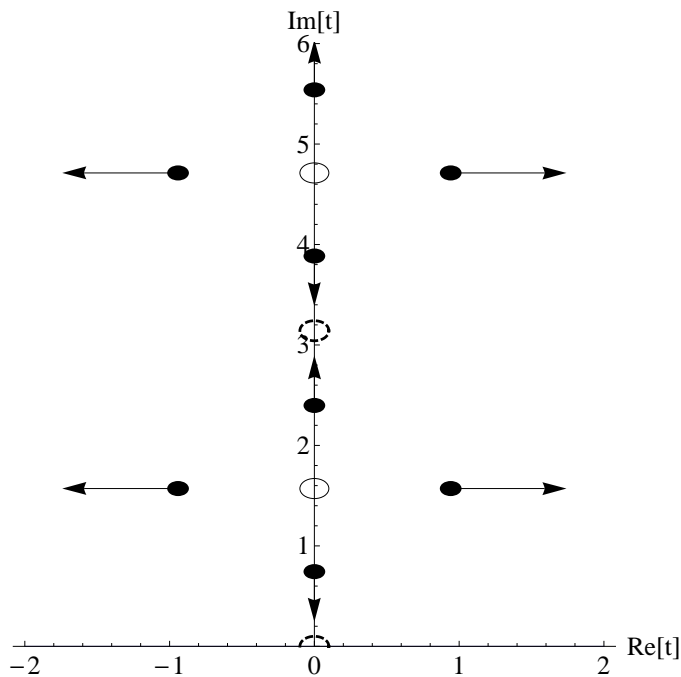} 
\caption{The structure of complex zero points for the Sech-Tanh model. The zero points for a specific finite value of $b/a$ are denoted by black dot (here we have taken $b/a = 0.8$). For the values $b/a = 0$ and $b/a \rightarrow \infty$, these points are denoted by a solid and dashed circles, respectively. The arrows indicate the behavior of the zero points as $b/a$ gets larger.}
\label{fig:zeropointsST}
\end{center}
\end{figure}

The structure and the parameter dependence of complex zeros is the key to understand the dynamics of the Sech-Tanh model and it can be explained as follows (see also figure \ref{fig:zeropointsST}). When $b/a = 0$ we have $X_{\pm} = 0$ and therefore the argument $2X_{\pm} - 1$ starts from the value $-1$ for both $X_{+}$ and $X_{-}$. For $X_{+}$, the argument obtains values in the interval $\left[-1, 1\right]$ and tends monotonously to positive unity as $b/a \rightarrow \infty$. Therefore, the inverse hyperbolic cosine is in this case purely imaginary and the imaginary part tends to zero as $b/a$ increases. For $X_{+}$ and for each integer $k$ there is two zero points corresponding to the two signs of arcosh in (\ref{eqn:complextransitionpoint}) on equal distances from $\dot{\imath} k $. For positive (negative) $k$ these lie on the upper (lower) half of the complex plane. For $k = 0$ the other is on the upper plane, other on the lower, and these are also the zero points closest to the real axis. 

The minus argument, $2X_{-} - 1$, starts from the negative unity, as mentioned above, and tends to minus infinity as $b/a$ increases. Therefore, the arcosh has constant imaginary part and a positive real part that gets larger as $b/a$ grows. For $k = 0$, the positive sign of arcosh in the RHS of equation (\ref{eqn:complextransitionpoint}) gives positive real and imaginary parts for the transition points while negative sign gives them both negative. However, the value of imaginary parts are $\pm\frac{\pi}{2}$, so taking all the values of $k$ into account, the whole "k-ladder" leads to structure where the zero points are symmetrically with respect to both imaginary and real axes. 

Concluding, the zero points, infinite in number, divide into two classes corresponding to $X_{+}$ and $X_{-}$ that move parallel to imaginary and real axes, respectively, as the parameters change. Also, for non-zero value of $b/a$, there is a unique zero point, namely one corresponding to $X_{+}$ and $k = 0$ on the upper half plane that gives the dominant contribution in the DDP formula. 

\subsection{Comparison between numerical and DDP results}

Figure \ref{fig:sechtanhmodel1d} shows the behavior of the final transition probability $P$ for Sech-Tanh model for small and intermediate values of $aT$ and $bT$. For the resonance case, $a = 0$, $P$ is strictly zero due to the area theorem. For small values of the detuning, $P$ is oscillatory and this can be seen as a consequence of the interference effects arising from the double-peaked nature of the excitation pulse. It is also worthwhile to notice that the frequency of these oscillations is more or less independent of both $a$ and $b$. Furthermore, as is the case for positive asymmetric pulses, the zeros of $P$ are eliminated and it will tend to non-zero value for large $b$. Also, from the CPT mechanism explained in \ref{subsec:ST} it is clear that for large enough $a T$, $P$ will tend to unity for large $b/a$. Actually, as figure \ref{fig:sechtanhmodel1d} indicates, this happens already for $a T \approx 1$.

\begin{figure}[h]
\begin{center}
\includegraphics[scale=0.55]{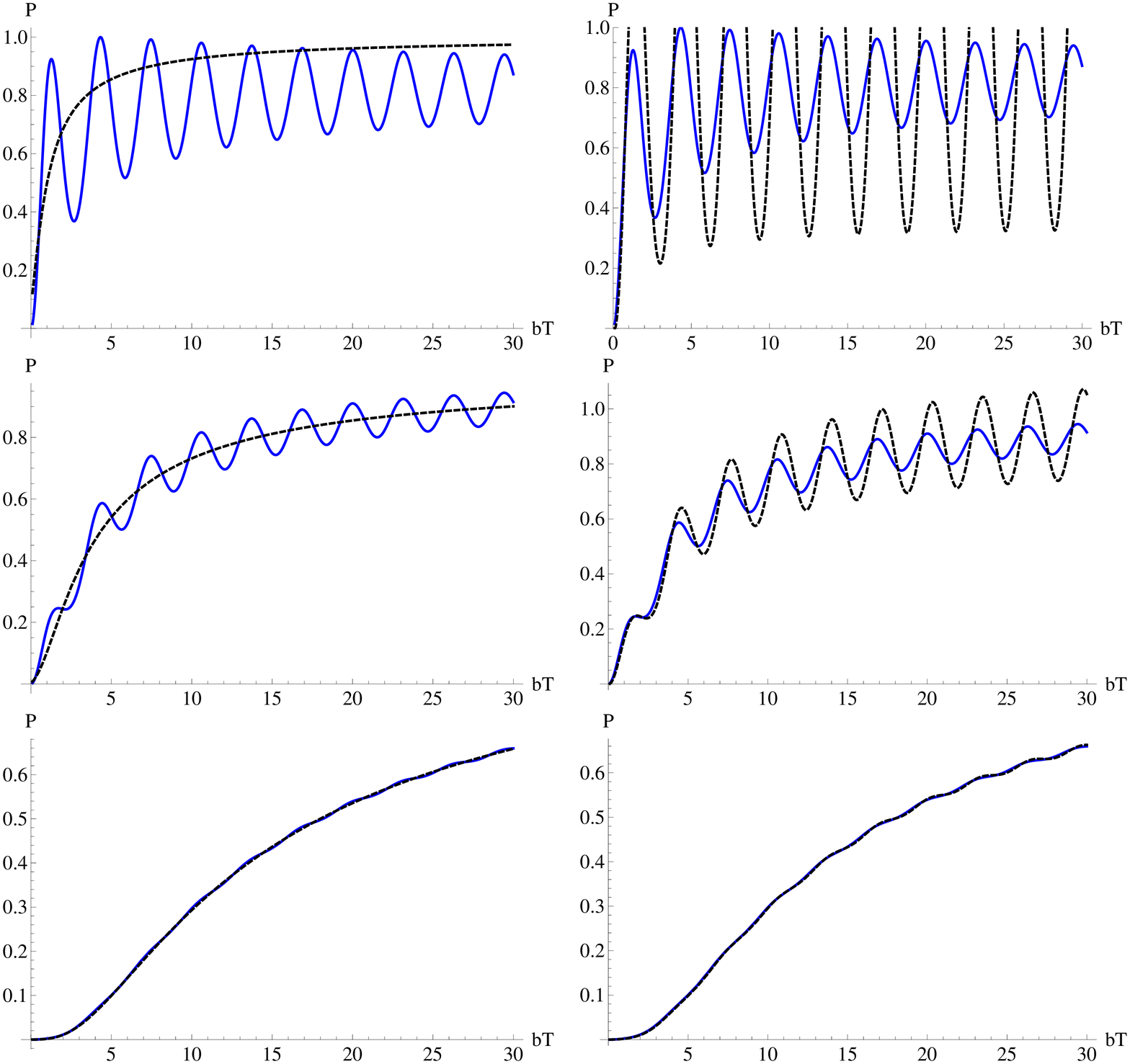} 
\caption{Here we have plotted the numerical exact (blue, solid line) and DDP (black and dashed line) solutions for $P$ for three different values of $a$:  in the uppermost plots we have $aT = 0.5$, in the middle $aT = 1$ and the lowermost plot is for $aT = 2$. In the left column, only the contribution from the first zero point is included and the dashed line is approximately given by the equation (\ref{eqn:approx1}). The right column shows the effect of including the  two zero points next nearest to real axis in the upper complex plane.} 
\label{fig:DDP1}
\end{center}
\end{figure}

In figure \ref{fig:DDP1} we illustrate the application of the DDP approach, taking into account the nearest zero points, to the Sech-Tanh model. The DDP phases (\ref{eqn:ddpintegral}) are solved numerically. Figure \ref{fig:DDP1} shows that by including only the nearest zero, corresponding to $X_{+}$ and $k = 0$, the DDP approach is able to give well enough approximation for $P$ in a large parameter region. It is clear that including only one term, corresponding to approximation (\ref{eqn:ddp}), cannot describe oscillations at all. However, the oscillations are very mild or nonexistent for $aT > 1 $ and then this formula provides a very good approximation indeed. For $aT = 1$, the formula (\ref{eqn:ddp}) is already meaningful, following the average behavior of $P$ and for $aT = 2$ it practically overlaps with the exact solution. It should be also noted that for very small values $b$, DDP is not concurrent with the exact solution and in that region the first Born approximation, $P = 4(aT)^{2}(bT)^{2}\pi^{2}\mathrm{sech}^{2}(\pi a T)$, is more adequate.

The qualitative features of the nearest zero point DDP approximation can be understood by studying the phase integral,
\begin{eqnarray}
D(\tau_{c}) &= 2 T \int_{0}^{\tau_{c}}\rmd x \sqrt{a^{2} + b^{2}\mathrm{sech}^{2}(x)\mathrm{tanh}^{2}(x)}\\ \nonumber
&= 2 a T\rmi \int_{0}^{\vert \tau_{c} \vert}\rmd x \sqrt{1 - (b/a)^{2}\sec^{2}(x)\tan^{2}(x)},
\label{eqn:STphaseintegral}
\end{eqnarray}

For the nearest zero point we have $\vert \tau_{c} \vert = \pi/2$ for $b/a = 0$ and it becomes smaller as $b/a$ becomes larger (see figure \ref{fig:zeropointsST}). Furthermore, also the integrand in the second line of (\ref{eqn:STphaseintegral}) behaves monotonously, decreasing from the value $1$ to $0$ in the integration interval. In the crudest approximation, the integrand is replaced by unity. Actually, a small modification to that formula, given by 
\begin{equation}
P \approx \exp \left[- \pi a T \mathrm{Im}\left(\mathrm{arcosh}(2X_{+} - 1)\right)/2 \right],
\label{eqn:approx1}
\end{equation}
is found to be in excellent agreement with the nearest zero point DDP approximation.

The oscillations in $P$ could be also thought to be a result from the interaction between the different zero points \cite{Vitanov1999}. When $b/a$ is not too large, the distances of the three nearest zero points from the real axis are comparable (see figure \ref{fig:zeropointsST}). The right column in figure \ref{fig:DDP1} shows what happens when we try to take the oscillations appearing in $P$ into account by including also the zero points corresponding to $X_{-}$ term. Although it is seen that this approximation can take some of the oscillatory character into account, giving some estimate for the frequency of the oscillation, for example, it is clearly not a very useful improvement, giving also values larger than one. This situation is not considerably altered by addition of further zero points. Unfortunately, both the zero point structure and the DDP phases are more complicated than for example in the case of Demkov-Kunike model and this prevents the straightforward summation of all zero points in the manner of \cite{Suominen1992c}. Therefore, it is presently an open problem whether including more zero points would increase the accuracy of the approximation, as in~\cite{LehtoSuominen2012}, or even produce the exact solution when the contribution from all the zero points is included, as happens for the Demkov-Kunike model \cite{Suominen1992a, Suominen1991, Suominen1992c}. 

\section{Discussion}

It is equally challenging to find analytic or even semi-analytic solutions for the zero pulse models as it is for any other two-level models. A tempting approach is to take a known model, create a dual model by swapping detuning and coupling, and then find a way to solve the dual model with the help of the original model. Therefore, a straightforward way to obtain exact solutions to models with zero pulse area would be to construct the propagator for a corresponding level-crossing model (i.e. with  $\varepsilon(t) = V_{ZPA}(t)$, $V(t) = const.$), if possible, and then to obtain the propagator for the model of interest by making a $\pi/4$ rotation, which has the effect of interchanging the detuning and Rabi frequency in the Hamiltonian. An example of such treatment can be found in~\cite{Torosov2008}. 

Unfortunately, such a reference model, to our knowledge, does not exists for the Sech-Tanh model. Indeed, there the Rabi frequency tends to zero in the initial and final times and such a time-dependence is not usually considered for the diabatic levels. Another possibility would be, after noticing their almost similar role in the equation, to interchange $V(t)$ and $\varepsilon(t)$ already in the two-level Schr\"{o}dinger equation. This would be very interesting in particular from the viewpoint of the DDP theory because the structure of the complex zero points would not be altered. It can be shown, however, by using the theory of invariants of ordinary differential equations, that the only models linked in this way are trivial~\cite{LehtoNotes}.

In any case, the DDP theory, applied here to the zero-area pulses, proves once again to be a useful approach to the problem even outside the adiabatic region for which it was originally devised. In particular, it can be used to approximate $P$ for already quite small values of $b/a$ where the simple estimate that was used to prove the existence of CPT, namely equation (\ref{eqn:pcpt}), does not work. One interesting feature is also that, unlike with the conventional models, such as the Landau-Zener model, the imaginary part of the complex transition point vanishes instead of the point receding from the real axis as the coupling gets larger. This elucidates the fact that the strong coupling limit for the Sech-Tanh model is different from the adiabatic limit.

In summary, we have reviewed and demonstrated the use and the properties of the DDP approach in solving time-dependent two-level models by applying it to a simple example of a zero area pulse model, namely the Sech-Tanh model. At the same time we have put forward the zero area pulses as a possible and so far neglected tool for quantum control, since by using off-resonant pulses, the self-cancelling nature of these pulses is eliminated, and even complete population transfer (CPT) can be obtained. Time-dependent two-level models have been an important tool in quantum physics and its applications over 80 years by now, and the fact that some basic and surprising effects such as the CPT mechanism are still being found, indicates that they continue to be a crucial core concept in modelling the interaction of matter with electromagnetic radiation in the future as well.

\end{document}